\documentclass[aps,pra,floatfix,twocolumn,groupedaddress,superscriptaddress,nofootinbib,notitlepage,amsmath,amssymb]{revtex4-1}
\usepackage{times,graphics,graphicx,bm,bbm,bbold,amsfonts,mathtools,dsfont,float,hyperref,amsthm}
\hypersetup{colorlinks,linkcolor={blue},citecolor={blue},urlcolor= {blue}}
\newtheorem{theorem}{Theorem}

\newtheorem{lemma}{Lemma}
\newtheorem{remark}{Remark}
\newtheorem{definition}{Definition}

\def\be{\begin{equation}}
\def\ee{\end{equation}}
\def\ba{\begin{eqnarray}}
\def\ea{\end{eqnarray}}
\def\la{\langle}
\def\ra{\rangle}

\def\h{\hskip 1cm}

\def\lo{\longrightarrow}

\def\m{\mu}

\def\bfe{\boldsymbol{e}}
\def\bfn{\boldsymbol{n}}
\def\bfr{\boldsymbol{r}}
\def\bfb{\boldsymbol{b}}
\def\bft{\boldsymbol{t}}
\def\bfd{\boldsymbol{d}}
\def\bfx{\boldsymbol{x}}
\def\bfy{\boldsymbol{y}}
\def\bfz{\boldsymbol{z}}
\def\bfu{\boldsymbol{u}}
\def\bfv{\boldsymbol{v}}

\begin{document}

\title{Quasi inversion of qubit channels}

\author{Vahid Karimipour }
\affiliation{Department of Physics, Sharif University of Technology, Tehran 14588, Iran}
\affiliation{Abdus Salam International Center for Theoretical Physics, Trieste, Italy}
\author{Fabio  Benatti}
\affiliation{Department of Physics, University of Trieste, I-34151 Trieste, Italy}
\affiliation{Istituto Nazionale di Fisica Nucleare (INFN), Sezione di Trieste, I-34151 Trieste, Italy}

\author{Roberto Floreanini}
\affiliation{Istituto Nazionale di Fisica Nucleare (INFN), Sezione di Trieste, I-34151 Trieste, Italy}

%\date{\today}

\begin{abstract}
Quantum operations, or quantum channels  cannot be inverted in general. An arbitrary state passing  through a quantum channel looses its fidelity with the input.    Given a quantum channel ${\cal E}$, we introduce the concept of its quasi-inverse as a map ${\cal E}^{qi}$ which when composed with ${\cal E}$ increases its average input-output fidelity in an optimal way. The channel ${\cal E}^{qi}$ comes as close as possible to the inverse of a quantum channel.   We give a complete classification of such maps for qubit channels and provide quite a few illustrative examples.   
\end{abstract}

\maketitle

%%%%%%%%%%%%%%%%%%%%%%%%%%%%%%%%%%%%%%%%%%%%%%%%%%%%%%%%%%%%%%%%

%\section{Introduction}
%\label{sec:intr}

Unitary dynamics of quantum systems is an idealization which almost never occur in reality. There are always inevitable and unknown couplings with the environment which destroy the coherence and purity of a quantum state and hence the information encoded into a quantum system. One of the central results in quantum theory is that a general non-unitary dynamics of an open quantum system can be characterized by operators acting entirely within the quantum system. This general dynamics is aptly called a quantum channel to signify the passage of quantum states (i.e. photons) through noisy environment (optical fibers or free air).  The most important goal of quantum communication is to combat this quantum noise which has led to whole subfields in quantum information science, like quantum error correction \cite{lam}, decoherence free subspaces \cite{lid,bei,kwi}, pre- and post-processing \cite{kat,sun,dav,kim} by weak measurements \cite{aha,pry}.
A quantum channel being completely specified by operators inside a system, raises the natural and highly important question if it can be inverted by some other set of operators, that is if we can invert a quantum channel and retrieve the input state in the same way that we do for unitary dynamics. If this inversion is possible, it can simply replace or at least complement other techniques for quantum state protection. 
It is however well known that quantum channels  cannot be inverted unless they are simple unitary channels of the form $\rho\lo U\rho U^\dagger$.  In this letter we ask to what extent we can come close to a complete inversion and introduce the concept of quasi-inversion of a quantum channel. We formulate this question in precise form, solve it for the important case of qubit channels, classify the solutions and present several examples. \\

Given a quantum channel $\cal E$, its overall performance can be measured through
the average input-output fidelity
\be\label{fid}
\overline{F}({\cal E}):=\int d\phi\, \la \phi|{\cal E}\big(|\phi\ra\la \phi|\big)|\phi\ra\ ,
\ee
where the integral is taken over all input states. The measure of the integral is  taken to be unitary-invariant, {\it i.e.} $d\psi=d\phi$ for $|\psi\ra=U|\phi\ra$,  and normalized to $\int d\phi=1$.  We now ask if it is possible to perform a quantum operation at the output, which increases this average fidelity independently of the input state and in an optimal way: 
\begin{definition}
\label{def}
Let ${\cal E}:\rho\lo {\cal E}(\rho)$ be a quantum channel, i.e. a completely positive trace preserving map \cite{cho,kra}. Its quasi-inverse, denoted by  ${\cal E}^{qi}$, is any channel fulfilling the following two conditions: 
\be\label{def2}
\overline{F}({{\cal E}}^{qi}\circ {\cal E}) \geq \overline{F}({\cal E})\ ,\h 
\overline{F}({{\cal E}}^{qi}\circ {\cal E})\geq \overline{F}({\cal E}'\circ {\cal E}), 
\ee
where ${\cal E}'$ is any other channel. 
\end{definition}
\noindent 

\noindent
In this paper we will restrict our study to qubit channels which will be shown to have already a rather rich structure. We will prove that the quasi-inverse of a qubit channel can always be taken to be a unitary map ${\cal E}^{qi}(\rho)=V\rho\, V^{\dagger}$,  and that it is both a left and a right quasi-inverse. We then show how it can be determined explicitly and illustrate the method by examples. To this end we use two complementary ways for characterizing a qubit channel:\\
{\bf a) The Kraus representation} 
\be
\rho\lo {\cal E}(\rho)=\sum_i K_i\rho K_i^\dagger,
\ee
with
 $K_i=a_i+\bfb_i\cdot \boldmath{\sigma}$, where the  trace preserving condition $ \sum_i K_i^\dagger K_i=I$ imposes the constraints 
\be
\label{Prules3}
\langle a^*\,a\rangle\,+\,\langle\bfb^*\cdot\, \bfb\rangle=1\ ,\quad
\langle a\,\bfb^*\ra\,+\la\,a^*\,\bfb\rangle\,+\,i\langle\bfb^*\times\bfb\rangle=0.
\ee
Here we have introduced the shorthand notations $\langle c\rangle=\sum_i c_i$, $\langle\bfd\rangle=\sum_i\bfd_i$,  \\
{\bf b) The affine map:} 
\be
\bfr\lo \quad \bfr'=M\bfr\,+\,\bft\ ,
\ee
which the channel induces on the Bloch sphere. Here  $M=[M_{\alpha\beta}]$ is a real $3\times 3$ matrix and $\bft$ a vector in $\mathbb{R}^3$ with components 
($\alpha,\, \beta=1,2,3$):
\be
\label{Mt}
M_{\alpha\beta}=\frac{1}{2}{\rm Tr}\Big(\sigma_\alpha\, {\cal E}(\sigma_\beta)\Big)\ ,\quad t_\alpha=\frac{1}{2}{\rm Tr}\Big(\sigma_\alpha\, {\cal E}(\mathbb{I})\Big)
\ .
\ee
Note that for unital channels, {\it i.e.} those obeying ${\cal E}(\mathbb{I})=\mathbb{I}$, one has ${\bft}=\,0$.
\noindent Any qubit channel ${\cal E}$ can be decomposed in the canonical form \cite{rus}:
\be
{\cal E}={\cal U}\circ {\cal E}_c\circ {\cal V},
\ee
or
\be\label{canonical-form}
{\cal E}(\rho)=U{\cal E}_c\big(V\rho\, V^{-1}\big)U^{-1},
\ee
where $U$ and $V$ are unitary matrices, and ${\cal E}_c$ is a channel with a diagonal $M$ matrix, 
$\Lambda_c={\rm diag}(\lambda_1,\lambda_2,\lambda_3)$.
Correspondingly, the $M$ matrix of $\cal E$ can be rewritten as
$
M=R_U \Lambda_c R_V\ ,
$
where $R_U$ and $R_V$ are $SO(3)$ representations of  $U$ and $V$.
\noindent  The parameters 
$\lambda_{1,2,3}$ are real and satisfy \cite{fuj, rus} $
|\lambda_{1,2,3}|\leq 1\ ,\ (1\pm\lambda_3)^2\geq (\lambda_1\pm\lambda_2)^2$ which 
 which constrain the vector $\boldmath(\lambda_1, \lambda_2,\lambda_3)$ to lie inside a tetrahedron (see supplementary material). When ${\bf t}\ne 0$, these conditions are necessary but not sufficient.  \\
\noindent The connection between the Kraus representation and the affine map is obtained through the equations (\ref{Mt}) which give
\be
{\bft}=\la a^*{\bfb}+a{\bfb}^*+i{\bfb}\times {\bfb}^*\ra\ ,
\ee
and \ \ $M=S+A$\ ,
where the real symmetric matrix $S=[S_{\alpha\beta}]$ is given by 
\be\label{s}
{S}_{\alpha\beta}= (1-2\la{\bfb}\cdot {\bfb}^*\ra)\delta_{\alpha\beta}+\la \bfb_\alpha \bfb_\beta^*+\bfb_\alpha^*\bfb_\beta\ra\ ,
\ee
and the real antisymmetric matrix $A=[A_{\alpha\beta}]$  by 
\be\label{A}
A_{\alpha\beta}=-\sum_{\gamma=1}^3\epsilon_{\alpha\beta\gamma}\bfv_\gamma, \ \ \ \ {\bfv}=i\la a^*{\bfb}-a{\bfb}^*\ra\ .
%A=\ \left(\begin{array}{ccc}0&-v_3&v_2\\ v_3&0&-v_1\\ -v_2&v_1&0\end{array}\right)
\ee
In this paper, we sometimes denote a quantum channel ${\cal E}$  %by its corresponding 
with affine map pair $(M,\bft)$ as ${\cal E}_{M,\bft}$ or simply by the  pair $(M,\bft)$ itself.
It should also be noted that while it is straightforward to obtain the affine map from  its Kraus representation, the converse is not easy at all. Moreover not every affine map corresponds to a quantum channel. \\

\noindent {\bf Average Fidelity:}  From the definition (\ref{fid}), and the fact that any pure state can be written as $|\phi\ra\la \phi|=\frac{1}{2}(1+{\bfn}\cdot \bm{\sigma})$, $||\bfn||=1$, we can find the average fidelity of a channel. The ingredients that we need are 
\be
\la \phi|K_i|\phi\ra=\frac{1}{2}{\rm Tr}\big[K_i \, (1+{\bfn}\cdot \bm{\sigma})\big]=a_i+{\bfb}_i\cdot {\bfn}\ ,
\ee
$
\int d{\bfn}\ {\bfn}=0\ $, and \ $ \int d{\bfn}\ n_\alpha\, n_\beta=\frac{1}{3}\delta_{\alpha\beta}\ .
$
The result is
 \be \label{fidab}
 \overline{F}({\cal E})=\la a^*a\ra+\frac{1}{3}\la\bfb\cdot\bfb^*\ra\ ,
 \ee
 which in view of the trace-preserving property \eqref{Prules3} can also be written as
 \be\label{fidab2}
 \overline{F}({\cal E})=1-\frac{2}{3}\la\bfb\cdot\bfb^*\ra=\frac{1}{3}(1+2\la a^*a\ra)\ .
 \ee
We will see in the following that the matrix $B=[B_{\alpha\beta}]$, with
\be\label{bm}
B_{\alpha\beta}=\frac{1}{2}\big\la b_\alpha b_\beta^*+b_\alpha^* b_\beta\big\ra\ ,
\ee
plays a central role in determining the quasi-inverse of a channel. In terms of this matrix the average fidelity reads %can be written as 
\be\label{fidB}
\overline{F}({\cal E})=1-\frac{2}{3}\,{\rm Tr}(B)\ .
\ee
The average fidelity can also be determined from the affine transformation
\be
\label{fidM}
\overline{F}(\mathcal{E})=\frac{1}{2}\left(1+\frac{1}{3}\,{\rm Tr}(M)\right)\ .
\ee
 Note in passing that from (\ref {s}) and (\ref{bm}) ${\rm Tr}(M)=3-4\,{\rm Tr}(B)$ which implies the equality of the two expressions (\ref{fidM}) and (\ref{fidB}) for the average fidelity. Also note that when $M$ is symmetric, we can write  
\be\label{BMsym}
B=\frac{1}{4}\big[2M+\mathbb{I}-{\rm Tr}(M)\big]\ .
\ee
This relation will be important when we discuss the quasi-inverse of qubit channels with symmetric affine maps.  
We now state one of the main results of this letter: 
\begin{theorem}
\label{U} 
The quasi-inverse of any qubit channel can always be taken to be a unitary map.
\end{theorem}
The proof is detailed in supplementary material. It is important to note that the proof hinges upon a basic property  specific to qubit channels, namely any unital qubit channel is a random unitary channel \cite{aud}. It remains to be seen whether or not the quasi-inverse of a quantum channel in higher dimension can be chosen to be unitary.  \\

Given the canonical decomposition (\ref{canonical-form}), one may be tempted to relate the quasi-inverses of ${\cal E}$ and ${\cal E}_c$. The following  theorem and remark elaborate this point.
 \begin{theorem}\label{qith}
	The quasi inverse of the map ${\cal E}={\cal U}\circ {\cal E}_c\circ {\cal U}^{-1}$ is given by 
	$
	{\cal E}^{qi}={\cal U}\circ {{\cal E}_c}^{qi}\circ {\cal U}^{-1}\ .
	$	
	\end{theorem}
\begin{remark}
	It is by no means true that the quasi-inverse of a general channel  ${\cal E}={\cal U}\circ {\cal E}_c\circ {\cal V}$ is of the form ${\cal E}^{qi}={\cal V}^{-1}\circ {{\cal E}_c}^{qi}\circ {\cal U}^{-1}$. 
\end{remark}
\noindent The proof of  theorem \ref{qith}, together with comments concerning the remark can be found in the supplementary material.\\

\noindent {\bf Two classes of channels:}  It is now crucial to note from the relation $M=R_U\Lambda_c R_V$ that the affine matrix of a channel is symmetric if and only if it is of the form ${\cal E}={\cal U}\circ {\cal E}_c\circ {\cal U}^{-1}$ . This connection drastically differentiates between the quasi-inverse of qubit channels with symmetric  affine matrix (for which $U=V^{-1}$ in their canonical form) and qubit channels with non-symmetric affine matrix (for which $U\ne V^{-1}$). Interestingly the unitality of the channel, does not play any role in this distinction, except for the implicit role that the transition vector ${\bf t}$ plays in determining the range of the parameters $\lambda_i$ \cite{rus}. We will be more explicit on this in remark \ref{rem}.\\

\noindent {\bf Explicit form of the quasi-inverse:} To find the explicit form of this quasi-inverse, let the quasi-inverse be ${\cal E}^{qi}(\rho)=V\rho\, V^{\dagger}$. The average fidelity of the combined channel 
$$
{\cal E}^{qi}\circ {\cal E}=\sum_i (VK_i)\rho\, (VK_i)^\dagger
$$ 
can be obtained from (\ref{fidab2}). We simply need to determine the scalar coefficients of the new Kraus operators $VK_i=a'_i+{\bfb}'_i\cdot \boldmath{\sigma}$. Taking the unitary matrix  $V=x_0+i{\bfx}\cdot \bm{\sigma}$, with $x_0^2+{\bfx}\cdot{\bfx}=1$ we find  
\be\label{vk}
VK_i=(x_0+i{\bfx}\cdot \bm{\sigma})(a_i+{\bfb}_i\cdot \bm{\sigma})=a'_i+{\bfb'}_i\cdot\bm{\sigma}\ ,
\ee
where $
a'_i=x_0a_i+i{\bfx}\cdot{\bfb}_i\ .
$
Using (\ref{fidab2}), the fidelity of the combined channel is $\overline{F}=\frac{1}{3}(1+2\la a'^*a'\ra)$ which can be rewritten as  
\be
\label{unitarychan6a}
\overline{F}\big({\cal E}^{qi}\circ{\cal E}\big)=1-\frac{2}{3}\,{\rm Tr}(B)\,+\,\frac{2}{3}\, \bfx^T\cdot {\widehat B}\cdot \bfx
+\frac{2i}{3}\,x_0\ \la a^*{\bfb}-a{\bfb}^*\ra\cdot\bfx\ ,
\ee
where \be\label{btil} \widehat{B}\equiv B-\mathbb{I}+{\rm Tr}(B).\ee
By combining (\ref{btil}) and (\ref{BMsym}) we find
$\widehat{B}=\frac{1}{2}(M-{\rm Tr}(M)).$
Note that setting $x_0=1$ and ${\bfx}=0$ ($V=\mathbb{I}$), one gets back the fidelity of the original channel. Recalling the definition of the vector ${\bfv}$
in (\ref{A}) and also (\ref{fidB}), the increase of average fidelity 
$
\Delta\overline{F}({\cal E})\equiv\overline{F}({{\cal E}}^{qi}\circ {\cal E}) - \overline{F}({\cal E})
$
can then be written as:
\be\label{unitarychan6b}
\Delta \overline{F}({\cal E})=\frac{2}{3}\Big(\bfx^T\cdot\widehat{B}\cdot \bfx
+ x_0\,{\bfv}\cdot\bfx\Big)\ .
\ee
Maximizing its value over all unitary maps, {\it i.e.} maximizing over the real parameters $(x_0,{\bfx})$, subject to the constraint $x_0^2+{\bfx}\cdot {\bfx}=1$, determines the quasi-inverse of the channel. 
 It is convenient to rewrite the r.h.s. of (\ref{unitarychan6b}) in quadratic form:
\be\label{Q1}
\Delta\overline{F}({\cal E})= \frac{2}{3}\left(\begin{array}{cc}\!\!x_0&\!\!{\bfx}^T\end{array}\!\!\right)Q\left(\!\!\begin{array}{c}x_0\\ {\bfx}\end{array}\!\!\right)\ ,
\ee
where
\be\label{Qform}
Q=\frac{1}{2}\left(\begin{array}{cc}0&{\bfv}^T\\{\bfv}& 2\widehat{B}\end{array}\right)\ ;
\ee
its maximum value is given by:
\be\label{Q2}
\Delta\overline{F}( {\cal E})=\frac{2}{3}{\rm Max}\big(\lambda_{max},\, 0\big)\ ,
\ee
where $\lambda_{max}$ is the largest eigenvalue of the $4\times 4$ matrix ${Q}$. The normalized eigenstate $(x_0,{\bfx})^T$ corresponding to this largest eigenvalue will determine the quasi-inverse of $\cal E$, {\it i.e.} 
the unitary rotation $V=x_0+i{\bfx}\cdot {\bm\sigma}$, or equivalently 
$V=e^{i\phi\,\hat{
			\bfx}\cdot\bm{\sigma}}$, with $\cos\phi=x_0$ and ${\bfx}=\sin\phi\ \hat\bfx$, with $\hat\bfx$ the unit vector along $\bfx$. \\

A simple calculation from equation (\ref{vk}), shows that the value of $a'_i$ for both $VK_i$ and $K_iV$ are equal. This means that if we had sought a right quasi-inverse, we would have reached the same equations as in (\ref{Q1}) and (\ref{Q2}). 
This can also be seen from the affine map picture. Let ${\cal E}^{qi}$ and ${\cal E}$ induce respectively  the affine maps $(N,{\bft'})$ and $(M,{\bft})$. Then
\be\label{FFEE}
\overline{F}({\cal E}^{qi}\circ{\cal E})\equiv \frac{1}{2}(1+\frac{1}{3}{\rm Tr}(N\,M))
\ee
which is symmetric with respect to the interchange of the two channels. Therefore the quasi-inverse of a qubit channel is both a right and a left quasi-inverse. 
We now study further properties of quasi-inverses. 

\begin{theorem}
	\label{prop1}
	For all qubit channels $\mathcal{E}$ whose affine matrix is symmetric and positive, the quasi-inverse is the identity map, i.e. their average fidelity cannot be increased. 
\end{theorem}

\noindent
\begin{proof}
	A symmetric matrix is diagonalizable. Therefore in  a suitable basis it is in the form:
	\be\label{Mdiag}
	M\equiv \Lambda_c={\rm diag}\left( \lambda_1,\  \lambda_2,\ \lambda_3\right).
	\ee
In the same basis the matrix  $\widehat{B}$ is of the form:
	\be\label{Bdiag}
	\widehat{B}=-\frac{1}{2}{\rm diag} \left( \lambda_2+\lambda_3,\  \lambda_1+\lambda_3,\ \lambda_1+\lambda_2\right).
	\ee
		 which in view of (\ref{Q2}) implies that if all $\lambda_i$'s are non-negative, then $\Delta \overline{F}({\cal E})=0$. Therefore such a channel has a non-trivial quasi-inverse only if at least one of the eigenvalues of $M$, i.e. one of $\lambda_i$'s is negative. 
\end{proof}

\begin{remark}\label{rem}
	There is a basic difference between channels with symmetric and non-symmetric affine matrices. In the symmetric case, $\bfv=0$ and the eigenvectors of the matrix $Q$ in (\ref{Qform}) are of the form 	
	$(0,\hat{{\bfx}})^T$ with $\hat{{\bfx}}$ a unit vector. Therefore the quasi-inverse of such a channel, if different from identity, is  
	  an inversion (a $\pi$-rotation) around some axis, i.e. ${\cal E}^{qi}(\rho)=V\rho V^{\dagger}$, with $V=\hat{\bfx }\cdot \bm{\sigma}$, and $\hat{{\bfx}}$ a unit vector. In the non-symmetric case (\ref{A}), $\bfv\ne 0$ and the corresponding eigenvector will not necessarily have $x_0=0$ and hence the quasi-inverse will be a rotation with a specific angle depending on the channel parameters.
	\end{remark}
 Below we will present one example of each kind. More examples can be found in supplementary material. \\
 
{\bf Pauli Channel:} This is a channel with symmetric affine matrix. 
	 	 \be
	 {\cal E}(\rho)=p_0 \rho + p_1 \sigma_x\rho \sigma_x +p_2 \sigma_y\rho \sigma_y+ p_3 \sigma_z\rho \sigma_z\ .
	 \ee
	 with $p_i\geq0$ and $\sum_{i=0}^3p_i=1$. This leads to a diagonal $B$ matrix
	 \be
	 B={\rm diag}\left(p_1,\ p_2, \ p_3\right).
	 \ee
	 Its average fidelity is given by
	 \be\label{p0}
	 \overline{F}({\cal E})
	 \equiv 1-\frac{2}{3}{\rm Tr}(B)=\frac{1}{3}(1+2p_0)\ ,
	 \ee
	 After combining with the quasi-inverse, the increase of average fidelity is given by
	 \be\label{deltaf0}
	 \Delta \overline{F}({\cal E})=\frac{2}{3}\,\lambda_{max}=\frac{2}{3}(p_{max}-p_0)\ ,
	 \ee
	 where $p_{max}$ is the largest of the probabilities $p_1,p_2$ and $p_3$, so that
	 \be\label{pmax}
	 \overline{F}\big({\cal E}^{qi}\circ{\cal E}\big)=\frac{1}{3}(1+2p_{max})\ .
	 \ee
	 The quasi-inverse $V$ is now a reflection with respect to the axis corresponding to $p_{max}$ ({\it i.e.} the $x-$ axis if $p_1$ is the largest probability). Moreover, comparing (\ref{p0}) and (\ref{pmax}), we find that if 
	 $
	 p_0\leq \frac{1}{2}\h {\rm and} $ and $ p_{max}\geq\frac{1}{2}
	 $
	 then
	 $
	 \overline{F}({\cal E})\leq \frac{2}{3} $  and $ \overline{F}\big({\cal E}^{qi}\circ{\cal E} \big)\geq \frac{2}{3}\ .
	 $
	 This means that the quasi-inverse can indeed increase the average fidelity of a noisy channel which is below the value of $2/3$ corresponding to that of a ``classical'' random channel, to above this value. {\it Note that it is not always the case that the inversion is determined by one of the Kraus operators. An example where this is not the case is given in supplementary material.} 
	 
\begin{figure}[H]\centering\vspace{-0.8cm}
		\includegraphics[width=11cm,height=8cm,angle=0]{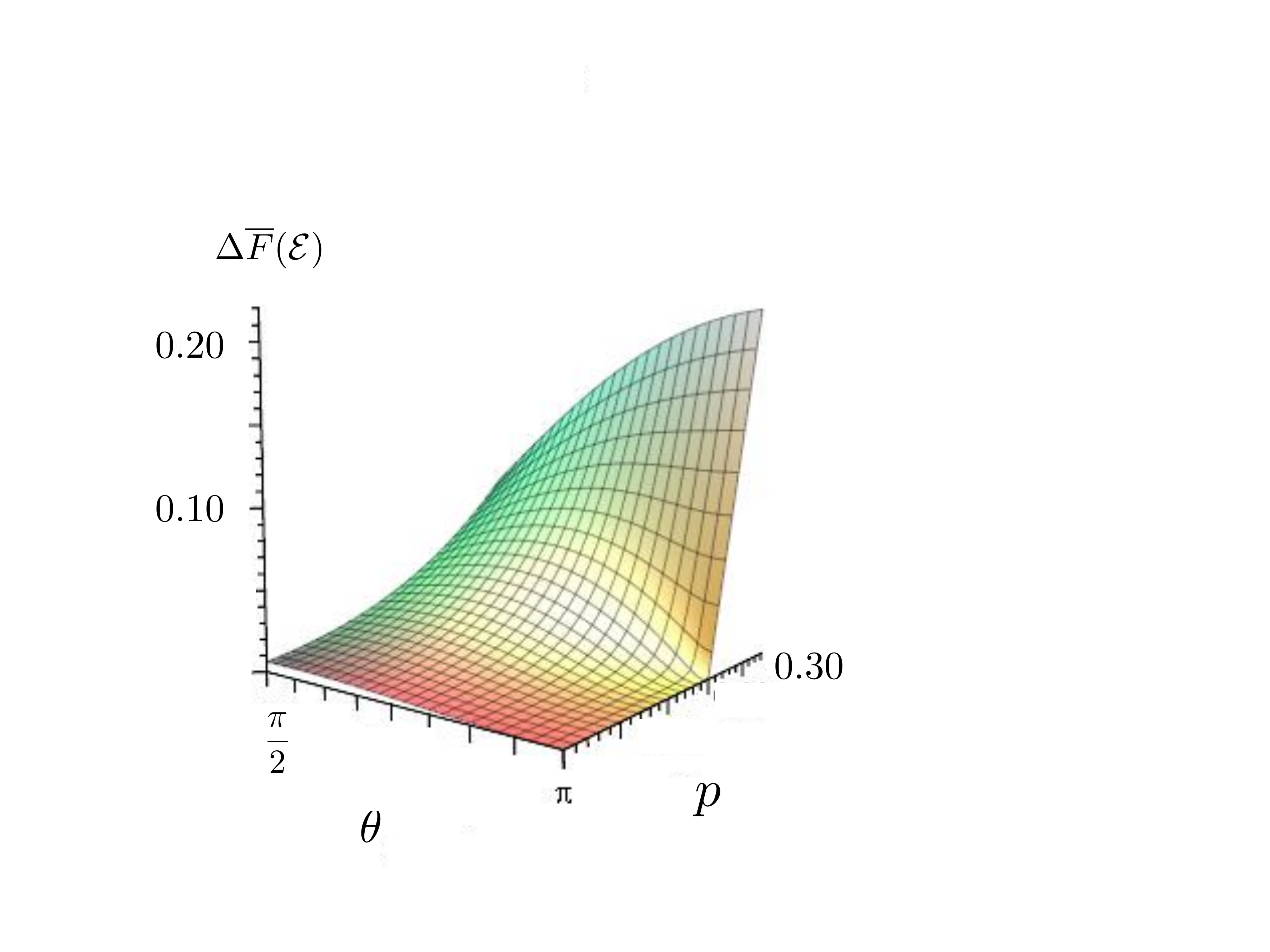}
		\vspace{-1cm}\caption{\small The increase of average fidelity for the random unitary channel given in (\ref{ru}).}\label{DeltaF}
	\end{figure}

	 {\bf A mixed unitary channel:} This is a channel with non-symmetric affine matrix. 
				\be\label{ru}
	{\cal E}(\rho)=p_0\rho + p\sum_{i=1}^3 U_i\rho\, U_i^\dagger ,
	\ee
	where
	$U_i=e^{-i\frac{\theta}{2} \sigma_i}$ is a rotation around the $\bfx_i$ axis with angle $\theta$ and $p_0+3p=1$.  The matrix $Q$ is now given by 
	\be
	Q=\left(\begin{array}{cccc}0 & v/2& v/2& v/2\\ v/2& q& 0 &0 \\ v/2& 0 & q&0 \\ v/2&0 &0 & q\end{array}\right)\ ,
	\ee
	where
	$
	v=p\sin \theta\ $ \  and $ \ q=4p\sin^2\frac{\theta}{2}-1\ .
	$
	For $q\geq0$, the largest eigenvalue of this matrix is 
	$
	\lambda_{max}=\frac{1}{2}(q+\sqrt{q^2+3v^2})\ ,
	$ 
	with corresponding eigenvector given by
	$
	|\lambda_{max}\ra\propto \left(\begin{array}{cccc}\frac{3v}{2\lambda_{max}}&1&1&1\end{array}\right)^T.
	$
	 This means that the quasi-inverse of the channel is given by the unitary $V=e^{i\phi\,{\bfn}\cdot {\bm{\sigma}}}$, where
	\be
	\cos \phi=\frac{\sqrt{3} v}{\sqrt{3v^2 +4\lambda^2_{max}}}\ ,\h {\bfn}=\frac{1}{\sqrt{3}}({\bfx}+{\bfy}+{\bfz})\ .
	\ee
	The increase in average fidelity is given by 
	$
	\Delta\overline{F}({\cal E})=\frac{2}{3}\, \lambda_{max}\ ,
	$
	which is plotted in Fig. \ref{DeltaF} as a function of parameters $p$ and $\theta$. 
	 
%\section{Outlook} 
%\label{sec:out} 
{\bf The problem of uniqueness:} It may happen that the matrix $Q$ of a channel ${\cal E}$ has two equal largest eigenvalue corresponding to two different unitary operators $V_1$ and $V_2$ as quasi-inverses, where $\overline{F}({\cal V}_1\circ {\cal E})=\overline{F}({\cal V}_2\circ {\cal E})$, leading to a one-parameter family of quas-inverses
${\cal E}^{qi}_p=(1-p){\cal V}_1+p{\cal V}_2$, not all members of which are unitary. Of course for generic channels the  quasi inverse is unique and unitary, since this degeneracy happens only for a set of measure zero in the space of all qubit channels.  A geometric elaboration of this point together with an explicit example is given in the supplementary material. \\

We have introduced the concept of quasi-inverse of quantum channels and have proved several of its properties for qubit channels, including their unitarity,  and equality of left and right inverses. A concrete formalism for finding the 
 quasi inverse of general qubit channels has been introduced and several classes of examples have been studied in detail. Let us  note that the quasi-inverse of a channel is different from its time reversal introduced in \cite{karol}
 in the context of entropy production in open quantum systems. The operation $R$  is not unitary and is an involution ${({\cal E}^R)}^R={\cal E}$, in direct contrast to the quasi-inverse.    This research raises several important questions, including the degree that the quasi-inverse of a channel can  partially restore the coherence of input states \cite{ple, mk, zan}, the extension of the concept and in particular the questions of  unitarity to higher dimension. While we have considered general qubit channels, the proof of the unitarity of the quasi-inverse hinges on a  very specific property of unital qubit  channels,  \cite{aud}, according to which any unital qubit channel is a random unitary channel. Counter example to this theorem in higher dimensions was  first found by Landau and Streater \cite{LS}. Therefore it would be quite interesting to see if there are quantum channels in higher dimensions whose quasi inverse are not unitary. One can also extend the notion of quasi-inverse  to the classical domain, i.e. to classical stochastic or bi-stochastic processes. 
%An analogy with classical random walks is interesting here: in a 1D random walk, it is enough to determine whether the random walker has on the average moved forward or backward and then one can make a deterministic walk to go back as close as possible to the original point under the given constraints. Is it possible to compensate in this sense a random walk in higher dimension by a deterministic walk or we should use another random walk.
 Finally the results presented have certainly practical importance for partial compensation of  noise in quantum channels; indeed, one may first apply the quasi-inverse  to increase the average input-output fidelity and then use error correcting techniques to completely recover quantum states transmitted through such channels. \\
 
\noindent
\textit{Acknowledgements.} 
We would like to thank S. Filippov, S. Raeisi, K. Zyczkowski and L. Memarzadeh for their  valuable comments and suggestions.  The work of V. K.  was partially supported by a grant no. 96011347 from the Iran National Science Foundation and the grant no. G950222 from  Sharif University of Technology.   F.B. and R.F. acknowledge that their research has been conducted within the framework of the Trieste Institute for Theoretical Quantum Technologies.

{}
\vspace{1cm}

\begin{center}
	\large{Supplementary Material for Quasi Inversion of Qubit channels}
\end{center}
{\bf A: Proof of Theorem 1:}
\begin{proof}
	First consider the definition of average fidelity. An important property of this quantity is its linearity, that follows from
	its definition in Eq. (1) in the main text, whence
	\be
	\overline{F}\Big(\sum_i\lambda_i{\cal E}_i\Big)=\sum_i \lambda_i \overline{F}({\cal E}_i)\ ,
	\ee
	where $\sum_i\lambda_i=1$ and $\lambda_i\geq 0 \,,\ \forall \lambda_i$.\\
	\noindent We now use a theorem of \cite{aud} according to which a necessary and sufficient condition for a qubit channel to be a random unitary channel, namely a convex combination of unitaries, is that it should be unital. Note that this theorem is not true for higher dimensions and holds only for qubit channels. Suppose now that the quasi-inverse ${\cal E}^{qi}$ is  unital. This means that ${\cal E}^{qi}=\sum_i p_i\, {\cal U}_i$, where ${\cal U}_i(\rho)=U_i\rho\, U_i^\dagger$ is a unitary map, and
	\be
	\overline{F}({\cal E}^{qi}\circ {\cal E})\geq \overline{F}( {\cal E})\ .
	\ee 
	%Then we use the above result and take ${\cal E}^{qi}=\sum_i p_i\, {\cal U}_i$, where ${\cal U}_i(\rho)=U_i\rho\, U_i^\dagger$ is a unitary map. 
	Therefore we have
	\be
	\overline{F}\Big[\sum_i p_i\, {\cal U}_i\circ {\cal E}\Big]\geq \overline{F}( {\cal E})\ .
	\ee
	
	\noindent Let ${\cal U}_{max}$ be the unitary map which has the highest contribution on the left hand side. Then it is obvious that if we replace all the random unitaries on the left hand side by ${\cal U}_{max}$, we get an even higher average fidelity:
	\be
	\overline{F}\big[{\cal U}_{max}\circ {\cal E}\big]\geq \overline{F}\Big[\sum_i p_i\,{\cal U}_i\circ {\cal E}\Big]= \overline{F}\big[{\cal E}^{qi}\circ {\cal E}\big]\geq \overline{F}( {\cal E})\ .
	\ee
	Therefore for any qubit channel whose quasi-inverse is unital, we can always take the quasi-inverse to be a simple unitary. 
	
	\noindent It now remains to see under what circumstances the quasi-inverse is unital. To solve this problem, it is useful to work with the channel represantation in terms of affine maps. Let $\Delta$ be the admissible domain of the parameters of the affine map defined by the pair $(M,\bft)$ and let ${\cal E}_{M,\bft}$ be the corresponding channel. Assume that ${\cal E}^{qi}_{N_0,\bft_0}$ be its quasi-inverse; according to Definition 1, this implies that
	\be\label{aa1}
	\overline{F}\big({\cal E}^{qi}_{N_0,\bft_0}\circ {\cal E}_{M,\bft}\big)\geq \overline{F}({\cal E}_{M,\bft})\ ,
	\ee
	and that for any other channel ${\cal E}_{N',\bft'}$, $(N',\bft')\in \Delta$, one has
	\be\label{bb1}
	\overline{F}\big({\cal E}_{N',\bft'}\circ{\cal E}_{M,t}\big)\leq \overline{F}\big({\cal E}^{qi}_{N_0,\bft_0}\circ {\cal E}_{M,\bft}\big)\  . 
	\ee
	In view of Eq. (21) these two conditions are equivalent to:  
	\be\label{a2}
	{\rm Tr}\big(N_0\,M\big)\geq {\rm Tr}(M)\ ,
	\ee
	and
	\be\label{b2}
	{\rm Tr}\big(N'\, M\big)\leq {\rm Tr}(N_0M)\ .
	\ee
	
	\noindent
	Note that although $\bft_0$ does not appear on the right hand side of this inequality, it affects the allowable range of $N_0$. 
	However, if ${\cal E}^{qi}_{N_0,\bft_0}$ is a CPT map, then ${\cal E}^{qi}_{N_0,0}$ is also a CPT map (the converse is not true, since the inclusion of the parameters ${\bft}$ restricts the allowable range of parametes of $M$ \cite{rus}). Therefore the average fidelity of the map 
	${\cal E}_{N_0,0}\circ{\cal E}_{M,\bft}$ is the same as the average fidelity of the map 
	${\cal E}_{N_0,\bft_0}\circ{\cal E}_{M,\bft}$ and both are determined by ${\rm Tr}(N_0\,M)$. 
	Thus, the two conditions (\ref{aa1}) and (\ref{bb1}) can be rewritten as:
	\be
	\overline{F}\big({\cal E}^{qi}_{N_0,0}\circ {\cal E}_{M,\bft}\big)\geq \overline{F}({\cal E}_{M,\bft})\ ,
	\ee
	and
	\be\label{b1}
	\overline{F}\big({\cal E}_{N',\bft'}\circ{\cal E}_{M,\bft}\big)\leq \overline{F}\big({\cal E}^{qi}_{N_0,0}\circ {\cal E}_{M,\bft}\big)\,,
	\h \forall  (N',\bft')\in \Delta\ . 
	\ee
	Therefore if the channel $(N_0,\bft_0)$ is the quasi-inverse for the channel $(M,\bft)$, then the unital channel $(N_0,0)$ is also the quasi-inverse for that channel with the same improvement of fidelity. This means that we can always take the quasi-inverse of a   qubit channel  to be unital and hence unitary according to the first part of the proof. \\
\end{proof}
{\bf B: Proof of Theorem 2}
We first need a lemma:
\begin{lemma}
	Let ${\cal E}_2$ and ${\cal E}_1$ be related as ${\cal E}_2={\cal U}\circ {\cal E}_1\circ {\cal U}^{-1}$, {\it i.e.} 
	\be
	{\cal E}_2(\rho)=U{\cal E}_1\big(U^{-1}\rho\, U\big)U^{-1}\ .
	\ee
	Then
	\be
	\overline{F}({\cal E}_2)=\overline{F}({\cal E}_1)\ .
	\ee
\end{lemma}\label{lem}
\begin{proof}
	The proof is straightforward once we use the definition of the average fidelity, make a change of variable $U|\phi\ra\lo |\psi\ra$ and use the invariance of the integration measure $d\phi=d\psi$.  
\end{proof}
We now turn to the main proof. 
\begin{proof}
	\noindent	From the above lemma, it immediately follows that if $${\cal E}={\cal U}\circ {\cal E}_c\circ {\cal U}^{-1}$$ then 
	$\overline{F}({\cal E})=\overline{F}({\cal E}_c)$.
	We now note that the definition of quasi-inverse for the channel ${\cal E}_c$ implies
	\be\label{qinn}
	\overline{F}({{\cal E}_c}^{qi}\circ {\cal E}_c)\geq \overline{F}_{{\cal E}_c}\ ,
	\ee
	and for all other channels $\mathcal{E}'$
	\be\label{Phi}
	\overline{F}\big(\mathcal{E}'\circ {\cal E}_c\big)\leq \overline{F}\big({{\cal E}_c}^{qi}\circ {\cal E}_c\big)\ .
	\ee
	Define 
	\be\label{qic}{\cal E}^{qi}:={\cal U}\circ {{\cal E}_c}^{qi}\circ {\cal U}^{-1}.\ee 
	Then one finds
	\begin{eqnarray}
	\nonumber
	{\cal E}^{qi}\circ {\cal E}&=&({\cal U}\circ {{\cal E}_c}^{qi}\circ {\cal U}^{-1})\circ({\cal U}\circ {\cal E}_c\circ {\cal U}^{-1})\\
	&=&{\cal U}\circ ({{\cal E}_c}^{qi}\circ {\cal E}_c)\circ {\cal U}^{-1}\ ,
	\label{channel-c}
	\end{eqnarray}
	and using the  Lemma \ref{lem} once more, one obtains:
	\be
	\overline{F}\big({\cal E}^{qi}\circ {\cal E}\big)=\overline{F}\big({{\cal E}_c}^{qi}\circ {\cal E}_c\big)\geq \overline{F}({\cal E}_c)=\overline{F}({\cal E})\ .
	\ee
	This proves that ${\cal E}^{qi}$ as in (\ref{qic}) increases the average fidelity of ${\cal E}$.
	Now let ${\cal E}'$ be any other channel. We have
	\begin{eqnarray}
	\nonumber
	\overline{F}({\cal E}'\circ {\cal E})&=&\overline{F}({\cal E}'\circ {\cal U}\circ {\cal E}_c\circ {\cal U}^{-1})\\
	&=&\overline{F}\Big({\cal U}\circ \big[{\cal U}^{-1}\circ {\cal E}'\circ {\cal U}\circ {\cal E}_c\big]\circ {\cal U}^{-1}\Big)\ ,
	\end{eqnarray}
	and using again the  Lemma \ref{lem}, we find
	\be
	\overline{F}({\cal E}'\circ {\cal E})=\overline{F}\Big( \big[{\cal U}^{-1}\circ {\cal E}'\circ {\cal U}\circ {\cal E}_c\big]\Big)=\overline{F}({\cal E}''\circ {\cal E}_c)\ ,
	\ee
	where ${\cal E}'':={\cal U}^{-1}\circ {\cal E}'\circ {\cal U}$.
	Using equation (\ref{Phi}), we have
	\be
	\overline{F}({\cal E}'\circ {\cal E})\leq \overline{F}({\cal E}_c^{qi}\circ {\cal E}_c)= \overline{F}({\cal E}^{qi}\circ {\cal E})\ ,
	\ee
	where (\ref{channel-c}) has also been used. \\ 
	\begin{remark}
		\noindent We should stress that for general channels of the form  ${\cal E}={\cal U}\circ {\cal E}_c\circ {\cal V}$, one cannot simply write the quasi inverse as ${\cal E}^{qi}={\cal V}^{-1}\circ {{\cal E}_c}^{qi}\circ {\cal U}^{-1}$. It is true that 
		\be
		{\cal E}^{qi}\circ {\cal E}={\cal V}^{-1}\circ {{\cal E}_c}^{qi}\circ {{\cal E}_c}\circ {\cal V},
		\ee
		and hence according to Lemma \ref{lem} and Eq. (\ref{qinn}) 
		\be
		\overline{F}({\cal E}^{qi}\circ {\cal E})=\overline{{{\cal E}_c}^{qi}\circ {{\cal E}_c}}\geq \overline{F}({\cal E}_c)
		\ee
		where in the inequality we have used the definition of quasi-inverse of the channel ${\cal E}_c$. However we can no longer use the equality of $\overline{F}({\cal E}_c)$ and $\overline{F}({\cal E})$, since this equality is not valid when  ${\cal U}$ and ${\cal V}^{-1}$ in the canonical decomposition of the channel are different. 
	\end{remark}
	\iffalse
	The fact that the affine matrix of ${\cal E}_c$ is diagonal, implies that the quasi-inverse of ${\cal E}_c$ is one of the Pauli matrices, which when conjugated with $U$ (i.e. $U\sigma_i U^{-1}$) leads to an inversion about an axis different from the ones
	along $\bfx$, $\bfy$ or $\bfz$ axes. This axis is determined by the eigenstate corresponding to the largest eigenvalue of the matrix $\widehat{B}$. Since according to Eq. (31) and Eq. (32) the ordering of the eigenvalues of $\widehat{B}$ and $M=\Lambda_c$ are identical, the axis is given by $\bfx$, $\bfy$ or $\bfz$ if the largest eigenvalue of $M=\Lambda_c$ is $\lambda_1$, $\lambda_2$ or $\lambda_3$ respectively. 
	\fi
\end{proof}

\noindent {\bf C: The geometric picture:}\\

\begin{figure}[H]\centering\vspace{-0.4cm}
	\includegraphics[width=7cm,height=5cm,angle=0]{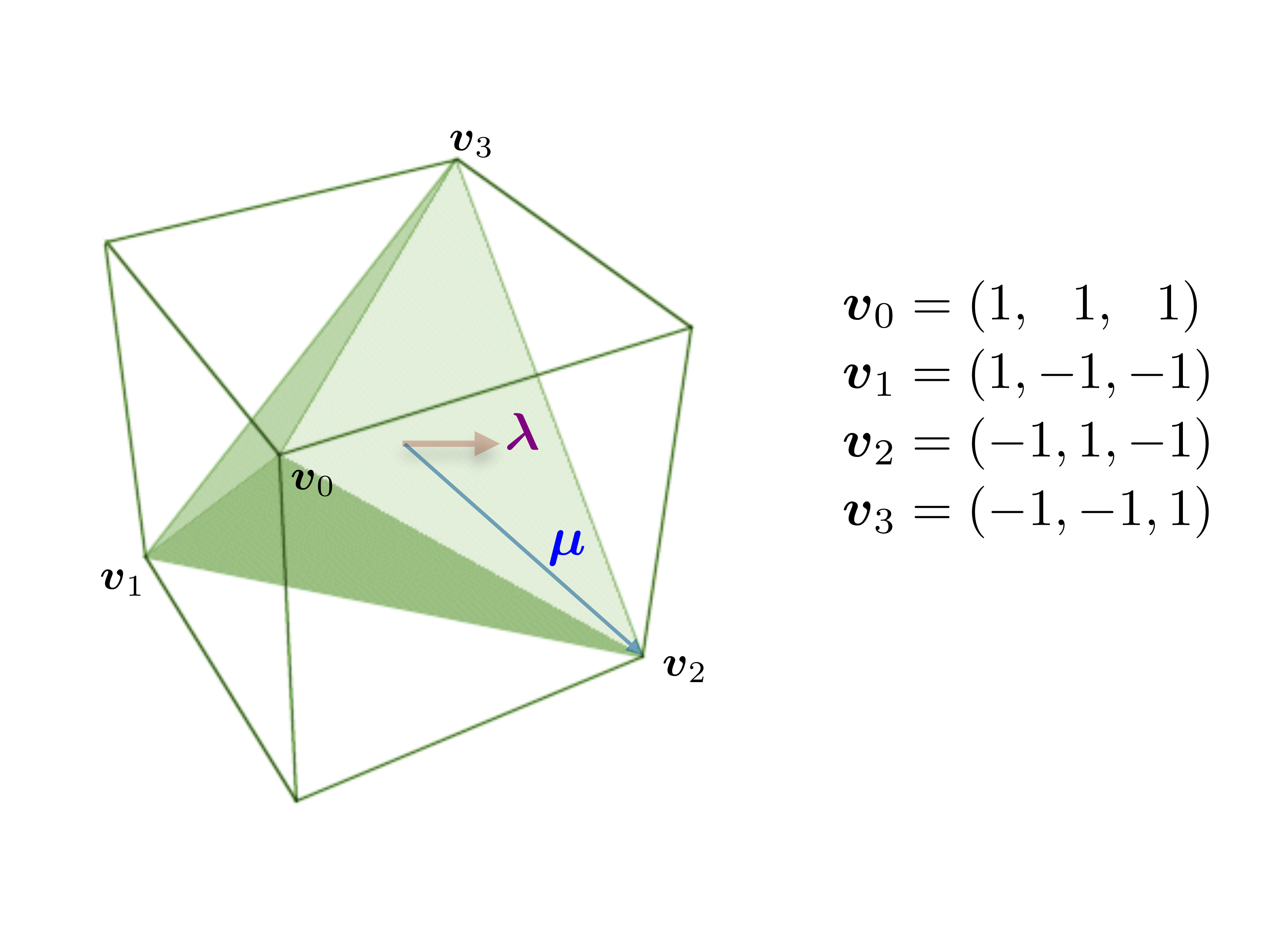}
	\vspace{-1cm}\caption{\small The canonical qubit channel ${\cal E}_c$ is characterized by the vector  ${\bm\lambda}$. Its quasi inverse ${\cal E}_c^{qi}$ is characterized by a vector $\bm{\mu}$ which maximizes the product ${\bm\lambda}\cdot {\bm\mu}$ in the expression $\frac{1}{2}(1+\frac{1}{3}\bm{\lambda}\cdot\bm{\mu})$.}\label{newtet2}
\end{figure}

One may  ask why we have not followed entirely the approach of affine maps for finding the quasi-inverse of a qubit channel by using equation
\be\label{FFEES}
\overline{F}({\cal E}^{qi}\circ{\cal E})\equiv \frac{1}{2}(1+\frac{1}{3}{\rm Tr}(N\,M))
\ee
and finding the matrix $N$ which maximizes the trace on the right hand side.
The problem is that even if one finds such a matrix by say numerical methods, it is not guaranteed that it defines a qubit channel. In fact while any qubit channel defines an affine map, not all affine maps define qubit channels. Nevertheless one can solve this problem for the special case of symmetric affine maps in a geometrical way. We note that such affine maps pertain to channels of the form  ${\cal E}={\cal U}\circ {\cal E}_c\circ {\cal U}^{-1}$, with ${\cal E}_c$ having a diagonal affine matrix, $\Lambda_c={\rm diag}(\lambda_1,\lambda_2,\lambda_3)$.
For complete positivity of the map (the qubit channel),  these parameters are confined to be inside a tetrahedron as shown in Fig.\ref{newtet2}. The corners of these tetrahedron correspond to ${\cal E}_i:\rho\lo \sigma_i\rho\, \sigma_i$ where $i\in \{0,1,2,3\} $ also includes the identity matrix $\sigma_0=\mathbb{I}$, that is to simple conjugation by Pauli matrices. The edges, faces and the inside of the tetrahedron correspond respectively to convex combination of two, three and four of these simple maps. According to Theorem 2, we only need to find the quasi-inverse ${\cal E}_c^{qi}$ whose affine matrix is again diagonal $N_c={\rm {diag}}(\m_1, \m_2, \m_3)$ with parameters in the same tetrahedron.
The parameters $\mu_i$ should be chosen to maximize \ the fidelity 
\begin{eqnarray}
\nonumber
\!\!\!\!\!\overline{F}\big({{\cal E}_c}^{qi}\circ {\cal E}_c\big)&=&\frac{1}{2}\bigg(1+\frac{1}{3}{\rm Tr} (N_c\Lambda_c)\bigg)\\
&=&\frac{1}{2}\bigg(1+\frac{1}{3}\big(\m_1\lambda_1+\m_2\lambda_2+\m_3\lambda_3\big)\bigg)\ .
\end{eqnarray}
If it were not for the constraint that the vector ${\bm {\m}}$  should be inside the tetrahedron, it could have been simply been taken   parallel to $\bm{\lambda}=(\lambda_1,\lambda_2,\lambda_3)$. However with this constraint and with our knowledge from Theorem 1 that the quasi-inverse can be a unitary map, 
it is enough to take the vector $\bm{\mu}$ to correspond to one of the vertices  $\bfv_0, \bfv_1, \bfv_2 $ or $\bfv_3$ depending on which one has the smallest Euclidean distance from $\bm{\lambda}$.  Inserting the coordinates of these vertices from Fig. \ref{newtet2}
into the following formulas 
\be\label{vvv}
||\bfv_0-{\bm\lambda}||\ , \quad
||\bfv_1-{\bm\lambda}||\ , \quad
||\bfv_2-{\bm\lambda}||\ , \quad
||\bfv_3-{\bm\lambda}||\ , 
\ee
and simplifying, we find that the comparison of these distances amounts to comparing the following expressions and determining which one is the maximum
\be\label{4lam}
\lambda_1+\lambda_2+\lambda_3\ , \lambda_1-\lambda_2-\lambda_3\ , \lambda_2-\lambda_1-\lambda_3\ , \lambda_3-\lambda_1-\lambda_2\ .
\ee
The maximality of these terms correspond respectively from left to right to the quasi-inverse being the identity operator or conjugation by $\sigma_1, \sigma_2$ and $\sigma_3$.
More concretely, when all the $\lambda_i$'s are non-negative, $\lambda_1+\lambda_2+\lambda_3$ is the largest of the above terms, which implies that $\bm{\lambda}$ is closest to $\boldmath{v_0}$ and hence the quasi-inverse is the identity map. When $\lambda_1\geq \lambda_2, \lambda_3$, then the second term in (\ref{4lam}) is the largest term and  $\bm{\lambda}$ is closest to $\bfv_1$ implying that the quasi-inverse is $\sigma_1$ etc.

\noindent Once the quasi-inverse of the canonical map ${\cal E}_c$ is obtained as one of the $\sigma_i$'s, the quasi-inverse of the map ${\cal E}$ is obtained from ${\cal E}^{qi}$ as $U\sigma_iU^{\dagger}={{\bfx}}\cdot {\bm{\sigma}}$ where ${\bfx}$ is the eigenvector corresponding to the largest eigenvalue of its matrix $M$  or $\widehat{B}$.\\

{\bf D: Uniqueness of the quasi-inverse} From the geometric picture we see that unless the tip of the affine  vector  $\boldsymbol{\lambda}$ is equi-distant to the corners of the tetrahedron, there is always a unique quasi-inverse which is a unitary (corresponding to the vertex closest to the tip of $\boldsymbol{\lambda} $). Only at this set of measure zero, we have degeneracy of quasi-inverses, where the convex combination of these quasi-inverses also leads to the same average fidelity and hence we have unital quasi inverses which are no longer pure unitary. As an  example, consider the channel ${\cal E}=\frac{1}{2}(\sigma_x\rho \sigma_x+\sigma_y\rho\sigma_y)$, corresponding to the middle of an edges of the tetrahedron, corresponding to the affine map $M={\rm diag}(0,0,-1)$. The quasi-inverse of this channel is ${\cal E}^{qi}_p=(1-p)\sigma_x\rho \sigma_x+p\sigma_y\rho\sigma_y$  for any $p$, leading to the channel 
$({\cal E}^{qi}_p\circ {\cal E})(\rho)=\frac{1}{2}(\rho+\sigma_z\rho \sigma_z)$ for which the affine matrix is $NM=\rm{diag}(0,0,1)$.\\

{\bf E: A channel whose quasi-inverse is different from one of its own Kraus operators}\\

\begin{figure}[H]\centering\vspace{-0.4cm}
	\includegraphics[width=9cm,height=7cm,angle=0]{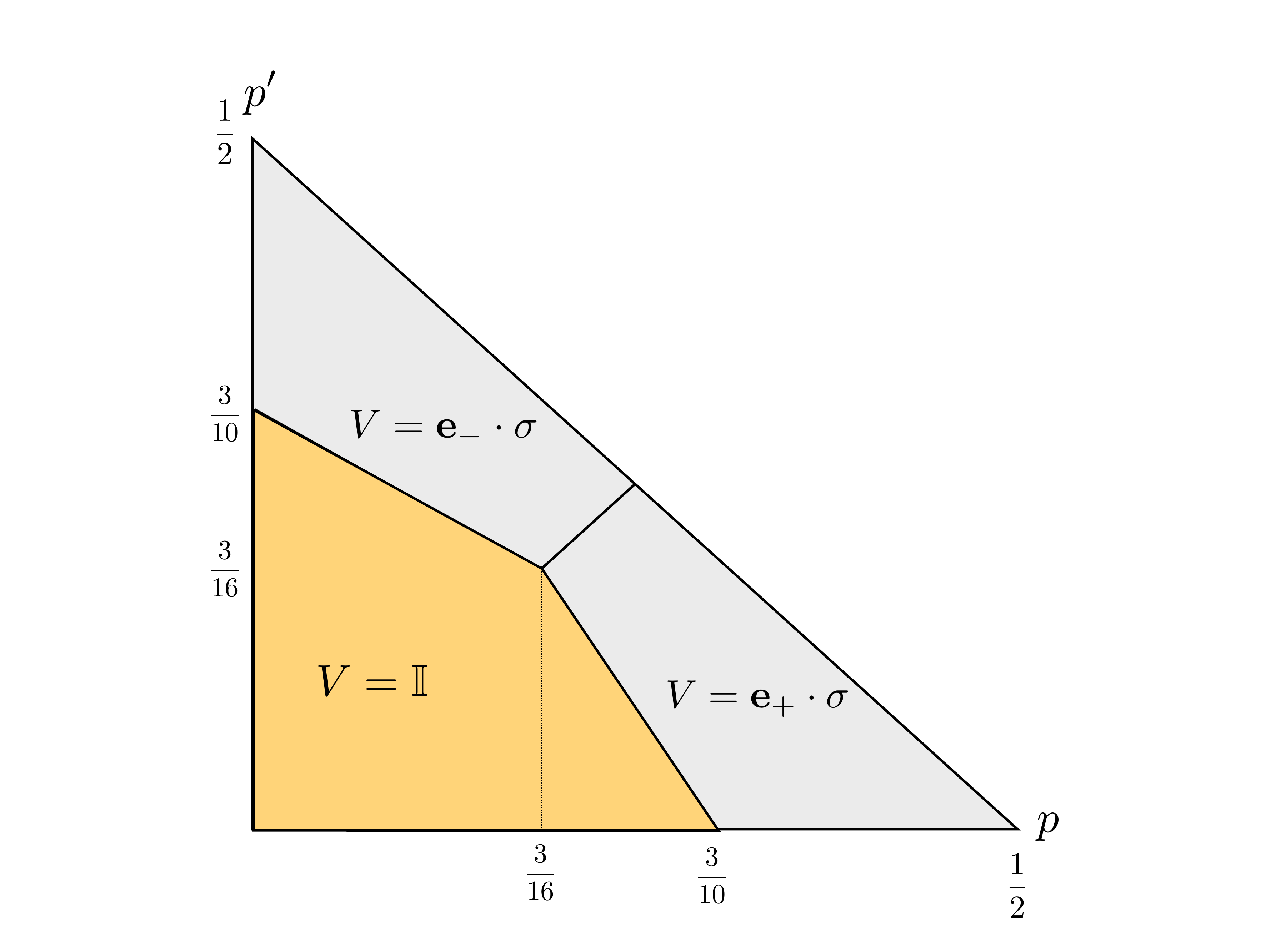}
	\caption{\small In the colored (yellow) region where $V = \mathbb{I}$, no increase is obtained in average fidelity. In the other (grey) regions the correcting unitary operator is shown. Here ${\bfe}_{\pm}=\frac{1}{2}({{\bfx}}\pm {{\bfy}})$ }\label{tet2}
\end{figure}

%\subsection{Tetrahedron Channel}
The Pauli channel is a special channel for which the quasi-inverse turns out to be one of Kraus operators of the channel, i.e. the Pauli matrices. There are many other channels for which this is not the case. In order to remain within the domain of analytical solutions and avoid numerical methods, we define a new channel and call it the tetrahedron channel.
%This seems to be a formal channel and not physically motivated. Its significance is to show that the correction 
%can be different from Pauli matrices and depending on the range of parameters, the quasi-inverse can change. 
The channel is defined by
\be
{\cal E}(\rho)=q\rho+ \sum_{i=0}^3p_i ({\bfu_i}\cdot \bm{\sigma})\,\rho\, ({\bfu_i}\cdot \bm{\sigma})\ ,
\ee
where $q=1-p_0-p_1-p_2-p_3$.  The vectors ${\bfu}_i$ are chosen to be the corners of a tetrahedron as
\ba
{\bfu_0}&=&\frac{1}{\sqrt{3}}(1,1,1)\ ,\cr
{\bfu_1}&=&\frac{1}{\sqrt{3}}(1,-1,-1)\ ,\cr
{\bfu_2}&=&\frac{1}{\sqrt{3}}(-1,1,-1)\ ,\cr
{\bfu_3}&=&\frac{1}{\sqrt{3}}(-1,-1,1)\ ,
\ea
so that the correspinding $\bfb$ vectors are given by $\bfb_i=\sqrt{p_i}\, \bfu_i$, $i=0,1,2,3$,
from which the corresponding $B$ matrix can be computed.
%It is seen that 
%\be
%\la {\bf b}_i\cdot {\bf b}_j\ra=\frac{-1}{3}\h i\ne j
%\ee
%The matrix $B$ is found to be 
%\be\label{b-matrix}
%B=\frac{1}{3}\left(\begin{array}{ccc}1-q&p_0-p_1-p_2+p_3&p_0-p_1+p_2-p_3\\ p_0-p_1-p_2+p_3& 1-q& p_0+p_1-p_2-p_3\\ p_0-p_1+p_2-p_3&p_0+p_1-p_2-p_3&1-q\end{array}\right)\ .
%\ee
For simplicity we consider the special case where  
\be
p_1=p_2=p, \h p_0=p_3=p',
\ee
with $p+p'\leq 1/2$ due to the normalization condition $q+2p+2p'=1$. With this choice, one finds:
\be
B=\frac{1}{3}\left(\begin{array}{ccc}2p+2p'&2p-2p'&0\\ 2p-2p'& 2p+2p'& 0\\ 0&0&2p+2p'\end{array}\right),
\ee
with eigenvalues 
\be
\lambda_1=\frac{4p}{3}\ ,\h \lambda_2=\frac{4p'}{3}\ , \h\lambda_3=\frac{2p+2p'}{3}\ ,
\ee
and corresponding eigenvectors
\be
{\bfe}_+=\frac{1}{\sqrt{2}}({\bfx}+{\bfy})\ ,\h {\bfe}_-=\frac{1}{\sqrt{2}}({\bfx}-{\bfy})\ ,\h {\bfe}_3={\bfz}\ , 
\ee
where $\bfx$, $\bfy$, $\bfz$ here denote the cartesian three dimensional unit vectors. The original average fidelity of this channel is given by 
\be
\overline{F}({\cal E})=1-\frac{2}{3}\,{\rm Tr}(B)=1-\frac{4}{3}(p+p')\ ,
\ee
and the increase in average fidelity is given by 
\be
\Delta\overline{F}({\cal E})=\frac{2}{3}\,{\rm Max}\Big[\lambda_{max},\  0\Big]\ ,
\ee
or explicitly
\be
\Delta\overline{F}({\cal E})=\Bigg\{\begin{array}{ccc} \!\!\!\frac{2}{3}\, {\rm Max} \{2p'-1+\frac{10p}{3} ,0\}&  {\rm if}\ p\geq p'\ ,\\ \!\!\!\frac{2}{3}\,{\rm Max} \{2p-1+\frac{10p'}{3} ,0\}& {\rm if}\ p\leq p'\ .\end{array}
\ee
The regions where an increase of fidelity is possible, and the unitary operator $V$ that achieves it, are shown in Fig.\ref{tet2}.\\

\noindent
Of course due to the symmetry of the Tetrahedron Channel, we can obtain, without further calculations, similar results if other pairs of probabilities are equal. 
%To this end, it is best to picture the vectors ${\bf b}_i$ as the vertices of a Tetrahedron, prescribed in a cube, as shown in figure (\ref{tet2}). 
When $p_0=p_2=p$ and $p_1=p_3=p'$, one finds the same results as before but with $\bfe_{\pm}=\frac{1}{\sqrt{2}}({{\bfx}}\pm {{\bfz}})$; similarly, the same holds when $p_0=p_1=p$ and $p_2=p_3=p'$ provided ${\bfe}_{\pm}=\frac{1}{\sqrt{2}}({{\bfy}}\pm {{\bfz}})$.\\ 

{\bf F: The amplitude damping channel}\\

%\subsection{The amplitude damping channel}\label{ad}
The amplitude damping channel ${\cal E}_{AD}$ is a non-unital characterized by the following Kraus operators
\be
A_0=\left(\begin{array}{cc}1&0\\ 0 &\gamma\end{array}\right)\ ,\h A_1=\left(\begin{array}{cc}0&\sqrt{1-\gamma^2}\\ 0& 0\end{array}\right)\ .
\ee
The Q matix is given by
\be
Q=\frac{1}{2}\left(\begin{array}{cccc}0&0&0&0\\ 0 & -\gamma(\gamma+1)&0&0 \\ 0&0&  -\gamma(\gamma+1)&0\\ 0&0&0&-2\gamma\end{array}\right)\ .
\ee 
As seen from above, for this channel, ${\bfv}=0$ and hence this is a channel with symmetric affine matrix. Later on we will consider a slightly twisted version of it which has a non-symmetric affine matrix. 
It is readily seen that if $\gamma>0$, then $\lambda_{max}$ is negative and hence no increase in average fidelity is possible , i.e. no non-trivial quasi-inverse exists. However for $\gamma<0$, the largest eigenvalue is $\lambda_{max}=-\gamma$ and $\Delta\overline{F}=-\frac{2}{3}\gamma$ implying that the quasi-inverse is $V=\sigma_z$. The fidelity of the channel itself is given from Eq. (18)  as $\overline{F}({\cal E}_{AD})=\frac{1}{2}+\frac{1}{6}\gamma^2+\frac{1}{3}\gamma$ and the fidelity of the combined channel is 
$\overline{F}({\cal E}^{qi}\circ{\cal E}_{AD})=\frac{1}{2}+\frac{1}{6}\gamma^2-\frac{1}{3}\gamma$. 
\iffalse Fig.\ref{ADChannel}  shows the fidelities of the two channel. \\
\begin{figure}[H]\centering\vspace{-0.6cm}
	\includegraphics[width=9cm,height=6cm,angle=0]{ADChannel}
	\vspace{-1cm}\caption{\small The standard amplitude damping channel with positive $\gamma$ has no quasi-inverse except the identity channel. Its average fidelity cannot be increased. However for negative values of $\gamma$, there is a quasi-inverse.  }\label{ADChannel}
\end{figure}
\fi
\begin{remark}
	If we denote the amplitude damping channel with negative $\gamma$, by ${\cal E}^-_{AD}$ and that with positive $\gamma$, which is the standard amplitude damping channel by ${\cal E}^+_{AD}$, then from the form of their Kraus operators, it is evident that ${\cal E}^-_{AD}={\cal Z}\circ {\cal E}^+_{AD},$ where 
	${\cal Z}\rho=\sigma_z\rho\, \sigma_z$. However from this relation, one cannot infer any conclusion between their quasi-inverses, since the concept of quasi-inverse as defined in this paper doesn't lead to a relation like 
	$(\Phi\circ {\cal E})^{qi}={\cal E}^{qi}\circ \Phi^{qi}$.
	\begin{figure}[H]\centering\vspace{-0.5cm}
		\includegraphics[width=8cm,height=6cm,angle=0]{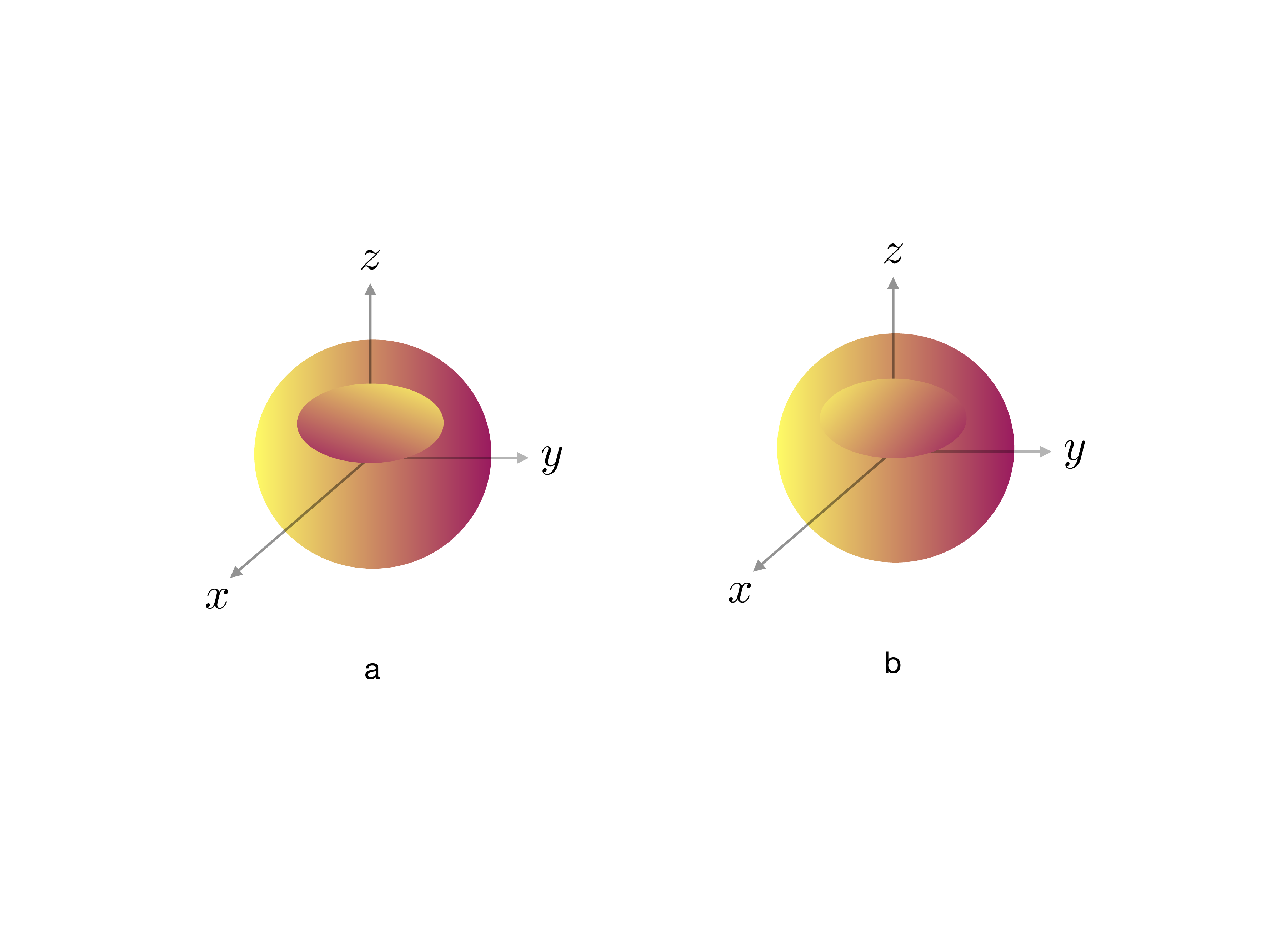}
		\vspace{-2cm}\caption{ \small The transformation of the Bloch sphere under the amplitude damping channels ${\cal E}^-_{AD}$ (a) and ${\cal E}^+_{AD}$ (b). }
	\end{figure}\label{2bloch}
\end{remark}
One may ask why a simple change of sign $\gamma\lo -\gamma$ makes so much difference in the quasi inverse of a channel. The answer is best seen when we look at the affine transformation associated to the amplitude damping channel: $M=\frac{1}{2}\rm{diag}(\gamma, \gamma, \gamma^2)$ and ${\bft}=(0,0,1-\gamma^2)^T$. When $\gamma>0$, the Bloch sphere is only shrunk and translated, but when $\gamma<0$, it is also reflected with respect to the $z$ axis (see Fig.4). The quasi-inverse compensates for this reflection and increases the average fidelity of the  $ {\cal E}^-_{AD}$, an action which if applied to  ${\cal E}^+_{AD}$ decreases the average fidelity instead of increasing it. \\

\noindent Consider now a slight modification of this channel when $A_0$ is changed to  
\be\label{ADcomplex}
A_0=\left(\begin{array}{cc}1&0\\ 0& i\gamma\end{array}\right).
\ee
The corresponding channel is still trace-preserving and non-unital but has a non-symmetric associated affine matrix. The matrix $Q$ is now given by
\be
Q=\frac{1}{2}\left(\begin{array}{cccc}0&0&0&\gamma \\ 0 & -\gamma^2&0&0 \\ 0&0&  -\gamma^2&0\\ \gamma&0&0&0\end{array}\right)\ ,
\ee 
so that $\lambda_{max}=\frac{|\gamma|}{2}$ and hence 
$\Delta\overline{F}=\frac{|\gamma|}{3}$. Then, the quasi inverse is the unitary $V=e^{i\frac{\pi}{4}\sigma_z}$. The average fidelity of this amplitude channel before applying the quasi inverse is 
\be
\overline{F}({\cal E_{AD}})=1-\frac{2}{3}{\rm Tr}(B)=\frac{1}{2}+\frac{1}{6}\gamma^2\ ,
\ee
and
\be
\overline{F}({\cal E}^{qi}\circ{\cal E_{AD}})=\frac{1}{2}+\frac{1}{6}\gamma^2+\frac{|\gamma|}{3}\ ,
\ee
which is larger than the average fidelity of the original channel for all values of $\gamma$. 

%\section{Channels with non-symmetric affine matrices}\label{nonsym}
{}

\end{document}